# Conceptual Modeling of Actions

Sabah Al-Fedaghi[*]

Computer Engineering Department
Kuwait University
Kuwait
salfedaghi@yahoo.com, sabah.alfedaghi@ku.edu.kw

*Abstract*—**Modeling in software engineering includes constructing static, dynamic, and behavioral representations. In describing system behavior, *actions* and *states* are two of the most commonly used concepts. In this paper, we focus on the notion of action. It generally held that the meaning of the concept of action is not that easy to grasp. According to some researchers, many existing systems do involve the notion of action, but in an obscure way. In Unified Modeling Language (UML), an action is a single atomic step within an activity, i.e., it is not further decomposed within the activity. Activity represents a behavior that is composed of actions. This paper contributes to the establishment of a broader interdisciplinary understanding of the notion of action in conceptual modeling based on a model called the thinging machine (TM). The TM uses only five primitive actions: create, process, release, transfer, and receive. The goal of such a venture is to improve the process of developing conceptual models by refining basic concepts such as action and event. To demonstrate how TM modeling represents actions, UML activity and Business Process Model and Notation (BPMN) diagrams are re-modeled in terms of the five TM actions. The results reveal the viability of the TM's five actions in modeling and relate them to other important notions such as activity, event, and behavior.**

## I. INTRODUCTION

Modeling in software engineering and systems engineering involves the process of collecting and analyzing system requirements to build a model that represents the domain (a part of the world) involved. The model includes constructing static, dynamic, and behavioral representations. In describing system behavior, actions, and states are commonly used concepts [1].

According to some scholars, modeling in its various forms is one of the most fundamental processes of the human mind since it allows us to see patterns, to appreciate and manipulate processes and things, and to express meaning [2]. Conceptual models are models consisting of concepts that shape the way we think about reality [3]. A conceptual model is an artifact that helps to understand a domain and, therefore, contribute to the elicitation of related functional requirements [4]. The resultant description plays a crucial role as a blueprint from which phases of development evolve. The conceptual picture describes a real-world domain while excluding technical aspects and serves as a guide for the subsequent design phase.

In software engineering, conceptual modeling is a crucial software development activity [5] that is situated in the broader view of requirements engineering [6].

According to González-Pérez [7] (see also [8]), conceptual modeling is usually associated with the discipline of software engineering, but modeling is a useful technique for exploring, documenting, understanding, and communicating artifacts of many kinds, and not only software. People in the humanities and social sciences do a lot of conceptual modeling. They may not use the same terminology that engineers do, but they still find conceptual modeling useful.

In this paper, we focus on modeling the notion of *action*. Action is of central interest to many disciplines, including economics, psychology, linguistics, law, business, and computer science. In philosophy, it has been studied since the beginning because of its importance for epistemology, and, in recent decades, it has even been studied for its own sake [9]. In computer science, the concept of action is present in many fields, e.g., artificial intelligence and in top-level ontologies like Cyc and SUMO and lexical resources like WordNet. In general, the action concept is present everywhere the dynamic aspects of the world are to be taken into account. According to Goldkuhl [10], the most fundamental property of a system in a business context is its actability, i.e., its ability to support actions and to perform actions.

However, "the meaning of the concept of action is not that easy to grasp" [11]. Many existing systems do involve the notion of action, but perhaps in an obscure way [10]. According to Zardai [12], "What is action?" is a question offered by philosophers in the last 80 years. Action is ambiguous and can mean either doing or the thing done [12]. In some domains, actions are confused with events [11]. Its structure seems to have a "complex and obscure nature" [13]. Various people have claimed that actions are events, that they are causings, and that actions are processes [12].

### A. What Is Action?

According to Trypuz [11], no current philosophical theories of actions (e.g., belief–desire–intention (BDI), STIT (sees to it that) logics) really answer the question of what an action is, and "in the context of building an ontology of action … [it] is hard conceptually and hard computationally." In philosophy of action, the word "action" is (in most cases) taken as being synonymous with the term *intentional action* [11]. For example, *Brutus killing Caesar* is an intentional action, because Brutus, when performing this action, had the intention to kill Caesar. For Searle [14], the whole action consists of two components—the intention-in-action and the bodily movement [11]. In general, the relation between the actions and the bodily movements is not so well established [11]. One sense of the term action is *something done*, usually as opposed to *something said*. *Something done* is a hyponym of an event ("something that happens at a given place and time") [11]. According to Goldkuhl [15], action means a change in the world, as the notion of action implies that an actor brings about some change, e.g., "when a firm manufactures goods, the employees are *acting* towards some material in order to create valuable products for customers" (italics added). Trypuz [11], going through the scientific fields, collected the following definitions of the action: (a) *action is an event*, (b) action is a bodily movement, (c) *action is an instance* of the relation of bringing about (or making happen), whose terms are agent and event, (d) action is the causing of an event, (e) action is a trying of an agent, (f) *action is a complex event*, (g) action is a bodily movement which is under the control of the agent, and (h) *action is a transition between states or situations*. The italicized phrases are highlighted because they are related to the TM notions of actions and events, where TMs are composed of five actions and events are instantiated actions.

The IGI Global Dictionary [16] provides the following definitions of action: (a) A fact, act, or operation that implies *activity*, movement, or change, (b) a process that uses the sources to obtain the task, (c) A sequence of goal-directed steps (d) An atomic and instantaneous unit of work done, (e) A complex sum of several movements, that is, the *sum of a discretizable event*, and (f) A descriptor of an actor's capability to perform operations. In contrast, as will be illustrated, in a TM, activities are composite elementary events. Kuutti [17] connects actions to actions. Activities can be considered to have three hierarchical levels: activity, action, and operation [mixed terminology, from the TM point of view]. A simple example of these levels is a description of the activity of

- *building a house* [in TM, composite actions that form events when injected with time], in which

- *fixing the roof* and *transporting bricks by truck* [in a TM, composite actions] are at the action level and

- *hammering* and *changing gears* when driving are at the operation level.

Traditionally, actions are called generic activities—that is, activities that cannot be divided into other activities. An activity is usually designated by a verb or verb form. Actions are identified and individuated in much the same way as entities [18].

### B. Actions in Unified Modeling Language

In Unified Modeling Language (UML) 2.0, an action is the fundamental unit of behavior specification. "An action takes a set of inputs and converts them into a set of outputs, though either or both sets may be empty. Actions are contained in activities, which provide their context. Activities provide control and data sequencing constraints among actions as well as nested structuring mechanisms for control and scope" [19]. According to Bock [20], in UML-2, the primitive actions (e.g., creating objects or invoking user-defined behaviors) "are not technically behaviors themselves, but this is more an artifact of metamodeling than a conceptual distinction".

The topic of UML action semantics is surrounded with difficult details. In this context, an action represents a computational procedure that changes the state of an element in the system [21]. Subclasses of primitive action are tailored to the kind of elements they act upon: null action, variable actions, object actions, and slot actions. There are two actions for objects: one to create a new object of a given class and another one to destroy an object. Compound actions are defined into group actions, conditional actions, and loop actions, each representing one of the basic language constructors. [21]. Alvarez et al. [21] give an example of the execution of some actions on an object as shown partially in Fig. 1. In this example, "the object is a dog with two slots, one indicating the length of its hair and a second one indicating its age. The birthday:WriteSlotExecution execution changes the value of the slot age of the object fido. The input of the execution is the value Three, that will be the value for the corresponding slot age of the next object in the history of fido" [21].

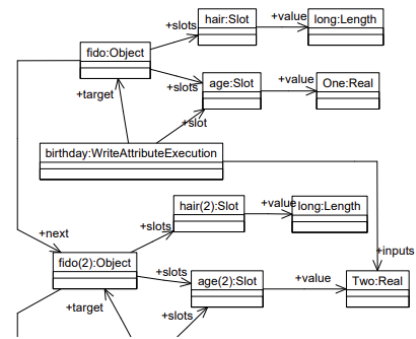

**Fig. 1. An example of the execution of actions on an object (partial)**

According to OMG.: UML 2.0 [19], because an action takes input and converts it to output, action seems to be a type of PROCESS as in the well-understood input-process-output model. The word PROCESS is capitalized to distinguish it from the action process, one of the generic TM actions. In UML diagrams [22], an action is a single atomic step within an activity, i.e., that is not further decomposed within the activity. An activity is defined as the coordination of elementary actions or it consists of one atomic action [23]. In UML, it is claimed that activity diagrams provide a high-level means of modeling dynamic system behavior. Aalst et al. [24] questioned the suitability of activity diagrams for modeling processes. According to Storrle and Hausmann [25], in UML "activity diagrams have always been poorly integrated, lacked expressiveness, and did not have an adequate semantics in UML." With version 2.0 of the UML activity diagram (AD), activities have been redesigned from scratch [23]. Several levels of increasing expressiveness were defined, including basic activities and activities by activity nodes.

According to UML diagrams [22], the name of the action is usually an action verb or a noun for the action, e.g., fill order, review document, checkout. An action may specify control flow and data flow. An action will not begin execution until all of its input conditions are satisfied. Action subclasses are: variable action: invocation action, raise exception action, structural feature action, link action, event action (not explicit in UML standard), and opaque action [22]. For example, object actions include different actions on objects, e.g. *create* and *destroy* object, *test* object identity, value, etc. Object actions are not defined explicitly by the UML standard [22]. Fig. 2 shows a sample event action.

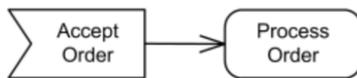

**Fig. 2. Example of event action (From [22])**

### C. *This paper*

This paper contributes to the establishment of a broader understanding of the notion of action in conceptual modeling based on a model called a thinging machine (TM) [26-27]. The TM model claims that there are five types of primitive action: create, process, release, transfer, and receive. The goal of such research is to improve the process of developing conceptual models by refining basic concepts.

Section 2 reviews the basics of TM modeling with some enhancements to the TM model. To demonstrate how TM modeling represents actions, section 3 models a classical example given by Sloman [28] that describes a mechanism that involves the actions of removing (stretchable) underpants without removing trousers. Sections 4 and 5 re-model UML activity and Business Process Model and Notation (BPMN) diagrams in terms of the five TM actions.

## II. THINGING MACHINE

The TM model is made of a composition of thimacs (*thing*s/*mac*hines), which corresponds to viewing domains as a complex of things that are simultaneously machines of five actions (see Fig. 3). A thimac is like a double-sided coin: one side of the coin exhibits the thing-ness, whereas, on the other side, actions emerge to form a machine. Modeling consists of a lower (static) structure of things that are simultaneously machines, and both merge into a thimac. At the upper level (dynamics), a time thimac combines with the static thimac to generate events structure. An analogy for a static thimac and a dynamic thimac is a running computer program. The program itself is a lifeless *thing*: it just sits there on the disk, instructions (actions) with maybe some static data [29]. Running a program is a dynamic thimac that is defined in terms of events, as will be discussed later.

The thimac is an encapsulation of as a thing that reflects the unity and hides the internal structure of the actions and a machine. Fig. 3 shows the structural components (called region), including *potential* actions of behavior. Behavior, in a TM, is a chronology of events that occur at particular time and over a region (static subdiagram). Potentiality, here, is a *degree* of reality that becomes full reality when actualized. In the TM static description, a thing is created at the level of potentiality, and that thing is *fully* actualized when it is event-ized (contains time). The potential reality (the static description) appears in the sense of *what is there*. If we take an instantaneous shot of, say, a business system, the static model consist of *what is there* in the system, a picture of a customer, enterprise, order, product, etc., connected by potential actions.

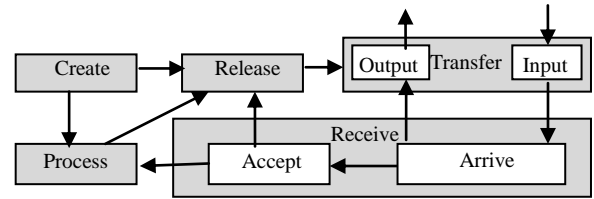

**Fig. 3. Thinging machine**

The fully manifested reality is realized when the thimacs in this picture are started as events in time, assuming that the content of the conceptual picture is reachable in reality. The event (as actions with a temporal extension) coincides with a certain change in the domain. The conceptual model can be viewed as a conjecture of reality and need not "look like" reality.

Things are what we start with when we construct the TM model. Each thing is defined in terms of at least the create (e.g., exist) action. *Create* produces new things in a machine, *process* manipulates things, *release* positions things before outputting them, *transfer* transforms things among machines and *receive* catches incoming things in a machine. A thing is subjected to *doing* (e.g., a tree is a thing that is planted, cut, etc.), and a machine *does* (e.g., a tree is a machine that absorbs carbon dioxide and uses sunlight to make oxygen). The tree thing and the tree machine are two faces of the tree thimac. The term thinging is taken from Heidegger's notion of thinging. A TM thing is clusters of potential generic actions that, with time, transform into atomic event constellations within specified acceptable behaviors.

A thimac is a thing. A thing is what can be created (seen, observed), processed (changed), released, transferred, and/or received. A thing is manifested (can be recognized as a unity) and related to the "sum total" of a thimac. The whole TM model occupies a conceptual "assemblage of actions" that forms a compositional structure of thimacs. The model whole is a grand thimac. The thimac "*being*" as a singular whole arises from the specific actions (including composite actions) and interactions of its constituent subthimacs. Thimacs can be "located" only via flow connections of actions among thimacs. A thing is a diagrammatic counterpart to a subdiagram of the TM static description (model). This subdiagram becomes an event (a thing with behavior) when merged with time.

The thimac is also a machine that creates, processes, releases, transfers, and/or receives. Fig. 3 shows a general picture of a (complete) machine. The figure indicates five "seeds" of potentialities of dynamism: creation, processing, releasing, transferring, and receiving.

TM modeling involves spatiality and actionality. The static description represents the (potential) actionality-based description that is expressed in terms of five generic actions. The static model and time are constructed in terms of these actions. As we will describe later, in TM modeling, actionality is a static notion that embeds the potentialities of events and behavior, which appear when time is added to the static model.

The generic/elementary (having no more primitive action) actions can be described as follows.

**Arrive**: A thing moves to a machine.

**Accept**: A thing enters the machine. For simplification, we assume that all arriving things are accepted; thus, we can combine the arrive and accept stages into one stage: the **receive** stage.

**Release**: A thing is ready for transfer outside the machine.

**Process**: A thing is changed, handled, and examined, but no new thing results.

**Create**: A new thing "comes into being" (is found/manifested) in the machine and is realized from the moment it arises (emergence) in a thimac. Things come into being in the model by "being found." While the action *create* indicates *there is such a thing*, when inside (a region) of an event, *create* indicates there is such a thing as of now.

**Transfer**: A thing is input into or output from a machine.

Additionally, the TM model includes the **triggering** mechanism (denoted by a ***dashed arrow*** in this article's figures), which initiates a (non-sequential) flow from one machine to another. Multiple machines can interact with each other through the movement of things or through triggering. Triggering is a transformation from movement of one thing to movement of a different thing.

## III. TM MODELING EXAMPLE: ACTIONS OF REMOVING UNDERPANTS WITHOUT REMOVING TROUSERS

Sloman [28] described a mechanism that involves actions of removing (stretchable) underpants without removing one's trousers (referring to the famous Mr. Been movie sketch). One solution given by Sloman [28] comprises "stretching the left side of the underpants down through the left trouser leg, over the foot and back up the left leg, leaving only the right leg through its hole. The underpants can then be slid down the right leg and out. A similar solution starts on the right side, with the underpants emerging through the left trouser leg." This series of actions provides an opportunity to model such sequence of actions diagrammatically in a TM.

Fig. 4 shows the static TM description of this solution. The left hole of the underpants (gray number 1) moves through the left side of the trousers (2 and 3). Then it moves to the area over the heel of the left foot (4), followed by the area over the toes (5), to arrive at the left leg of the trousers (6). Accordingly, the whole pair of underpants (7) moves through the right leg of the trousers (8) to the outside (9).

The *order* control of these series of actions is accomplished through the chronology of events. A TM event is a subgraph of the static TM model (called a *region* of the event) plus time. A region of an event is an action or a group of actions that form a TM subdiagram. When it is *active* (involves time), it becomes full event. For example, Fig. 5 shows the event *The left hole of the underpants moves to the left leg of the trousers*. For the sake of simplification, an event can be represented by its region.

Accordingly, we select the following events over composite actions (see the events model in Fig. 6).

E1: The left hole of the underpants moves to the left leg of the trousers.

E2: The left hole of the underpants moves from the left leg of the trousers to the area over the heel of the left foot.

E3: The left hole of the underpants moves from the area over the heel to the area over the toes of the left foot.

E4: The left hole of the underpants moves from the area over the toes to the area over the left leg of the trousers.

E5: The whole pair of underpants moves to the right leg of the trousers.

E6: The underpants move outside the right leg of the trousers.

Fig. 7 shows the behavior in terms of the chronology of events. This behavior is attributed to the system (trousers, underpants, and legs) and its control as a whole. The behavior model is a coordinated system of events that is formed to attain a certain function.

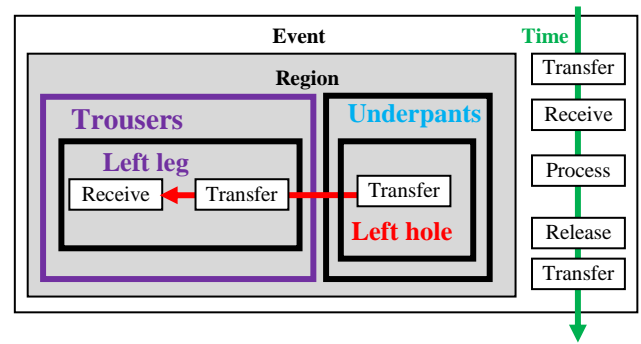

**Fig. 5. The event *The left hole of the underpants moves to the left leg of the trousers***

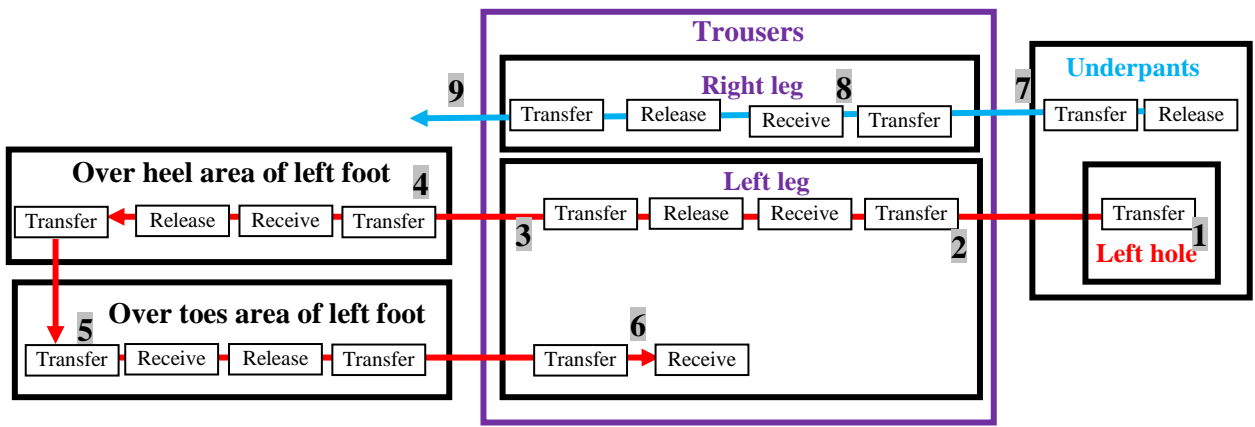

**Fig. 4. Static description of actions of removing underpants without removing one's trousers**

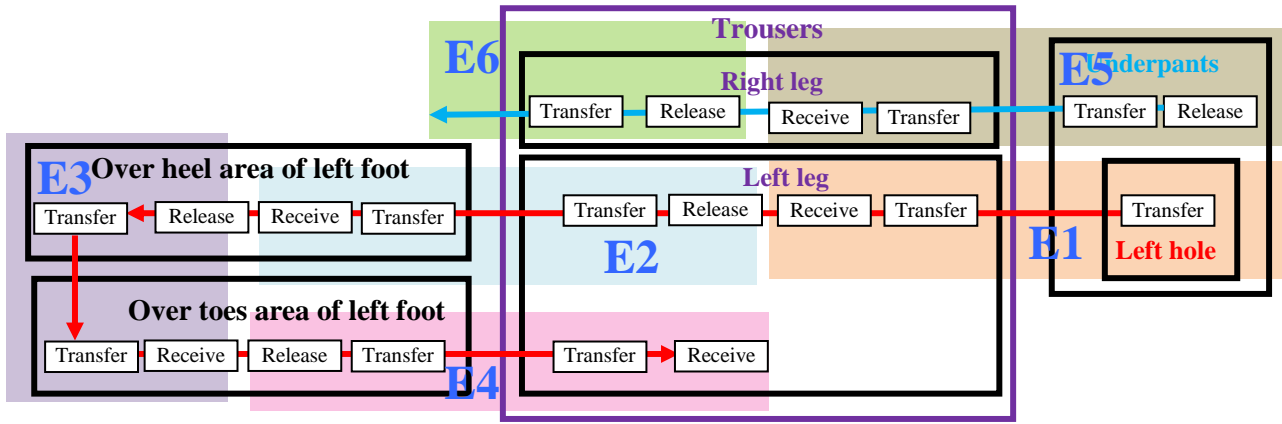

**Fig. 6. The events model**

$$E1 \rightarrow E2 \rightarrow E3 \rightarrow E4 \rightarrow E5 \rightarrow E6$$

**Fig. 7. The behavior model**

## IV. ACTIVITY Diagram and Actions

According to Bock [20], actions are the only elements in UML that have a persistent effect on objects, invoke operations on them, and invoke behaviors directly. For this reason, actions are sometimes called *primitive dynamic elements* in UML because all behaviors must eventually reduce to actions to have any effect on objects or even to invoke other behaviors. Additionally, according to Bock [30], in UML, a class supports the *behaviors* invoked by *actions*. Bock [30] provided the example shown in Fig. 8. An alternative notation is shown in the same figure (bottom), where partitions are labeled on nodes. The corresponding class model is shown in Fig. 9, where each company has instances of these classes. Fig. 9 could be used without the partitions before details are worked out. This facilitates UML application by modelers who do not use object orientation, such as system engineers and enterprise modelers.

Control flow edges in these figures connect actions to indicate that the action at the target end of the edge (the arrowhead) cannot start until the source action finishes. In this notation, there are two types of edges with different diagrammatic representation: *Control edges* connect actions directly, whereas *object flow edges* connect actions' inputs and outputs [30]. Actions' inputs and outputs, which are called *pins*, are connected by object flow edges to show how values flow through the activity, provided by some actions and received by others. [30].

### A. TM static model

Fig. 10 shows the TM static representation of the above example. An action in TM is defined in terms of the five elementary actions: create, process, release, transfer, or receive, as shown in the definition of *thinging machine*. First, in Fig. 10, an order is sent to Customer Service (pink number (1)). This triggers (a) the inauguration of the incoming order as an active order (2) and (b) the creation of an invoice that is sent to the customer (3). Additionally, the order is sent to Fulfillment (4), where it is processed (5) to trigger sending the product to the customer (6). After delivery, a delivery notification is sent to Customer Service (7).

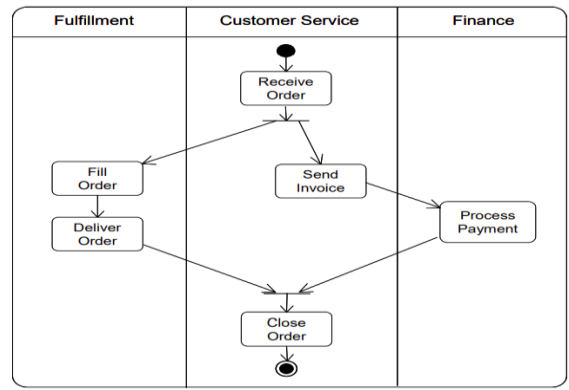

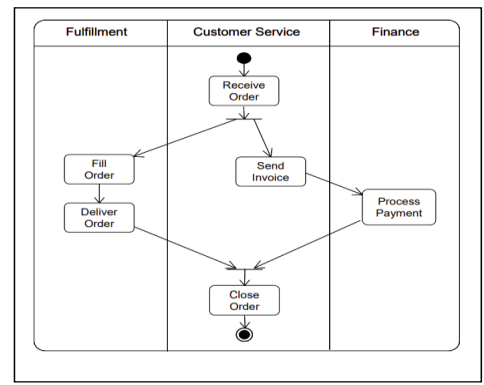

**Fig. 8. An example and an alternative notation (from [30])**

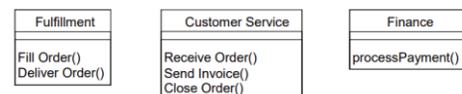

**Fig. 9. The UML class model (from [30])**

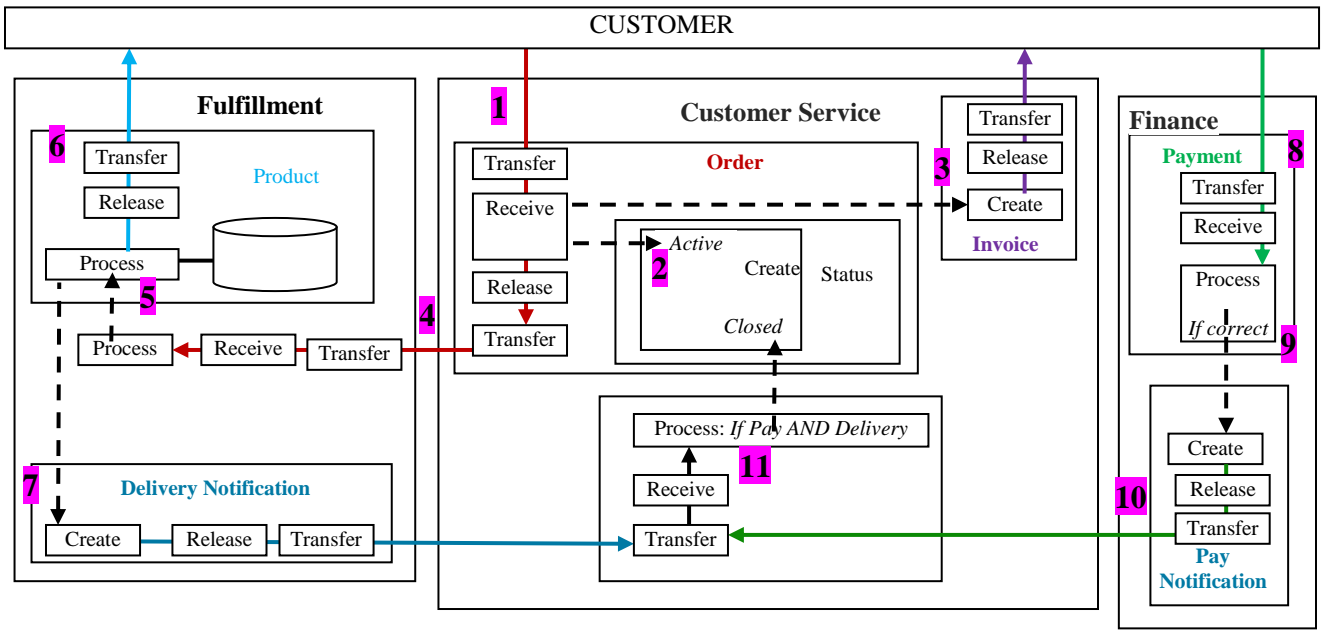

**Fig. 10. The static TM model**

When a payment is received from the customer (8), it is processed, and if it is correct (9), a payment notification is sent to Customer Service (10). In Customer Service, the payment and delivery notifications are processed, which triggers changing the order status to *closed* (11). Note that all arrows in this TM diagram denote a flow of things; a control flow will be specified in the behavior model, which will be developed next. Furthermore, note that Fig. 10 can easily be simplified at several levels. For example, we can assume that the arrows' direction indicate the direction of the flow of things; therefore, we can delete the actions, transfer, release, and receive, thus producing the simpler diagram.

The so-called class methods [30] are embedded into the TM representation. Fig. 11 identifies these methods of Customer Service. For simplification's sake, the actions are not shown under the assumption that the arrows indicate the direction of flow. In TM, Fig. 10 is a static representation that indicates potential actions. The dynamism is incorporated when time and static actions are encapsulated to form an event. A TM event is a region (a subgraph of the static description) plus time. Fig. 12 shows the event *Customer Service receives an order*. For simplicity's sake, an event may be represented by its region.

In TM modeling, we distinguish between static descriptions and dynamic activities. UML modeling mixes these aspects of modeling. Control flow edges in the activity diagrams connect actions to indicate that the action at the target end of the edge (the arrowhead) cannot start until the source action finishes.

As mentioned previously, in this notation, there are two types of edges with different diagrammatic representations (see Fig. 13); control edges connect actions directly, whereas *object flow edges* connect actions' inputs and outputs [30]. Fig. 14 illustrates why UML mixes the static and dynamic levels in Fig. 13.

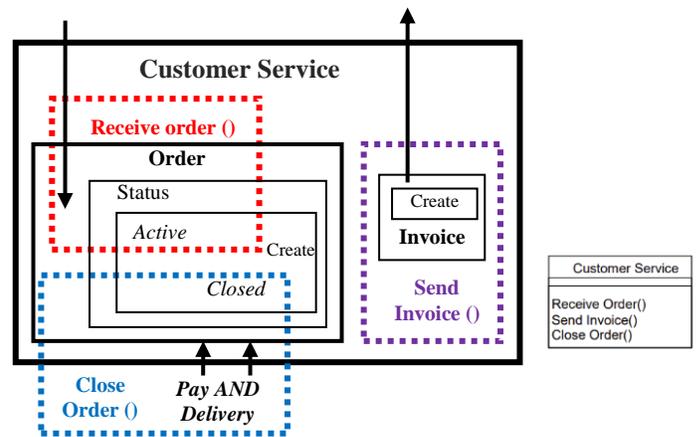

**Fig. 11. A subdiagram that corresponds to the Customer Service class**

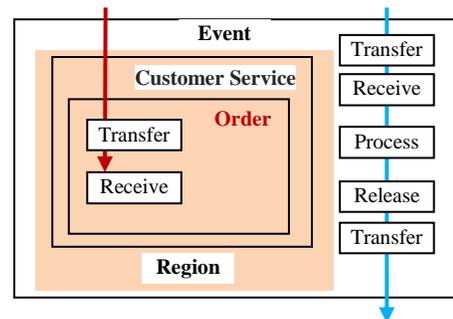

**Fig. 12. The event *Customer Service receives an order***

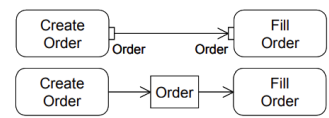

**Fig. 13. Example actions and object flow edges (from [30])**

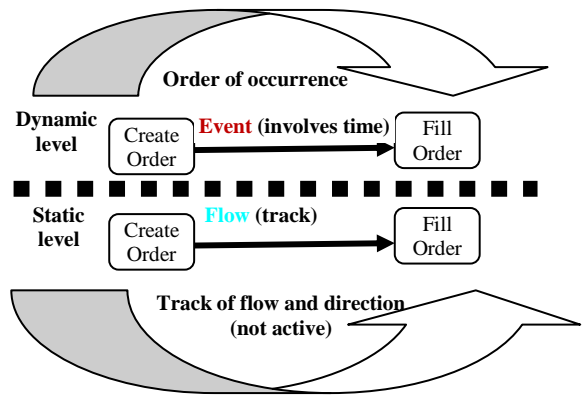

**Fig. 14. UML mixes static and dynamic levels**

## B. Events and behavior models

Accordingly, to develop the dynamics of the example under consideration, we can use the generic actions as regions of events. However, for clarity consideration, we can specify larger events. Fig. 15 shows the following selected events:
E1: Customer Service receives an order.
E2: The order is declared as an active order.
E3: An invoice is sent.
E4: The order is sent to Fulfillment.
E5: The product is sent to the customer.
E6: Fulfillment sends a notification to Customer Service.
E7: Payment is received.
E8: Finance sends a payment notification to Customer Service.
E9: The order is closed.

Fig. 16 shows the behavior model of the given example.

## IV. BPMN OR UML ACTIVITY DIAGRAM

Geambaşu [31] studied modeling business processes using the modeling languages BPMN and UML activity diagram (UML AD), which are the two graphical notations most used for the representation of business processes. BPMN can be described as a graph, the nodes of which are *actions* and the *arcs* of which are the flows between actions. The BPMN model can provide documentation and regulation functions and serve as an executable algorithm [32]. According to Stravinskiene and Serafinas [33], a business process is a complex phenomenon and is more than just a sequence of actions; it is "an agile, complex organizational unit with a logical and time-bound sequence of actions." UML AD covers aspects of process modeling, including actions, control flow, data and object flow, and the information related to process enactment. A behavior comprises a coordinated sequence of actions [34].

The given example involves a car service for damaged vehicles brought in by the company's customers (see Fig. 17). The process begins with a request made by the customer to the car service for vehicle repairs. Next, the car service schedules the repairs. When the repair start day comes, the customer brings the vehicle to the car service, which performs the required repairs.

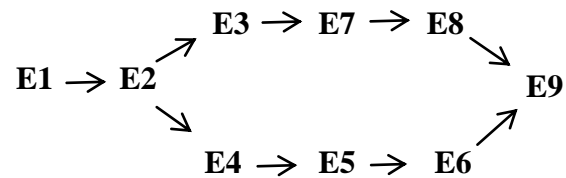

**Fig. 16. The behavior TM model**

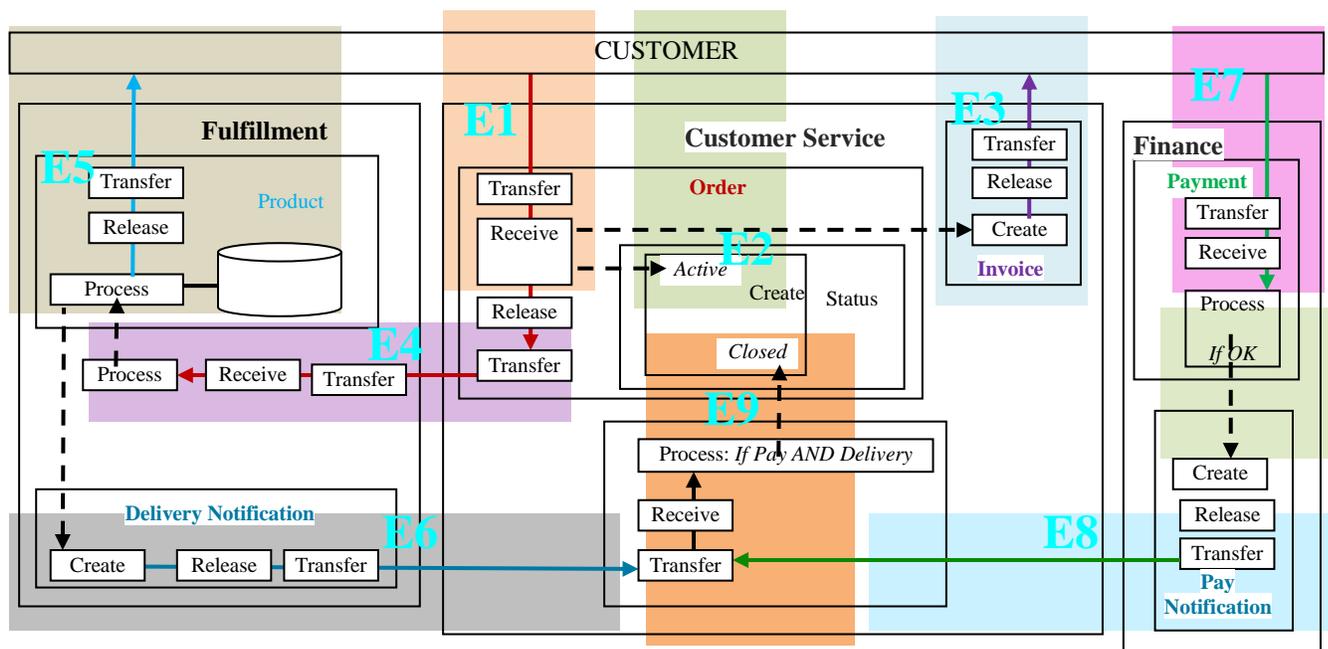

**Fig. 15. The events model**

When all the repairs are finished, the car service creates an invoice that must be paid by the customer to pick up their repaired vehicle (see Fig. 17) [31]. One objective is to analyze BPMN and UML AD considering how well the graphical elements of these two notation languages represent an organization's real business processes and how easily these two business process modeling languages can be mapped to business process execution languages.

The action node of UML AD corresponds to the task object in BPMN. BPMN uses a start *event* and an end event for the parts of the process corresponding to each participant, whereas UML AD uses only one initial node and one final node for the entire process. Activities (without loops) performed by the participants include the following: *Require vehicle reparation services*, *Schedule reparation*, *Bring damaged vehicle*, *Create invoice*, *Pay invoice*, and *Pick up vehicle from service*. Both representations mix static action and dynamic events (repair start date icons).

Fig. 18 shows the corresponding TM static model. First, the customer creates and sends a request for repairs (number 1) that is received and processed by the car service (2). Then, the request is input to a machine (3) that processes the request and its scheduling data (4) to trigger the creation of a repair appointment (5). According to this appointment, the car is delivered to the car device (6 and 7) to be processed (repeatedly, as will be modeled in the events model (8). At the end of this processing, an invoice is created (9) and sent to the customer (10). The customer processes (11) the invoice to trigger creating a payment (12) that flows to the car service (13). In the car service, the payment is processed (14); hence, the car is released to the customer (15 and 16).

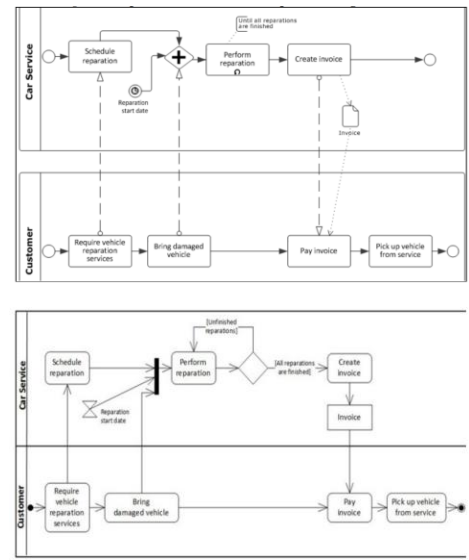

**Fig. 17. Representation of a business process using BPMN 2.0 (top) and UML AD (bottom)**

The events model is shown in Fig. 19 with the following events.

E1: The customer sends a request for car repairs that is received and processed by the car service.

E2: The request and the car service's scheduling data are processed.

E3: A car repair appointment is generated and sent to the customer.

E4: The car is delivered to the car service.

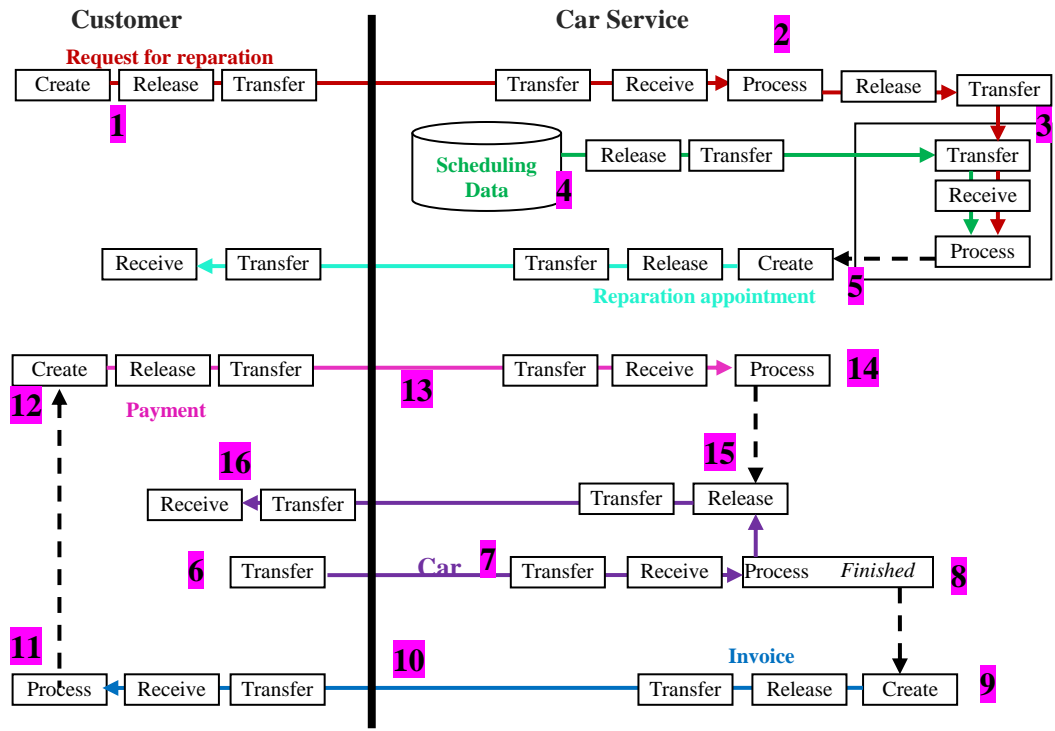

**Fig. 18. The TM static model**

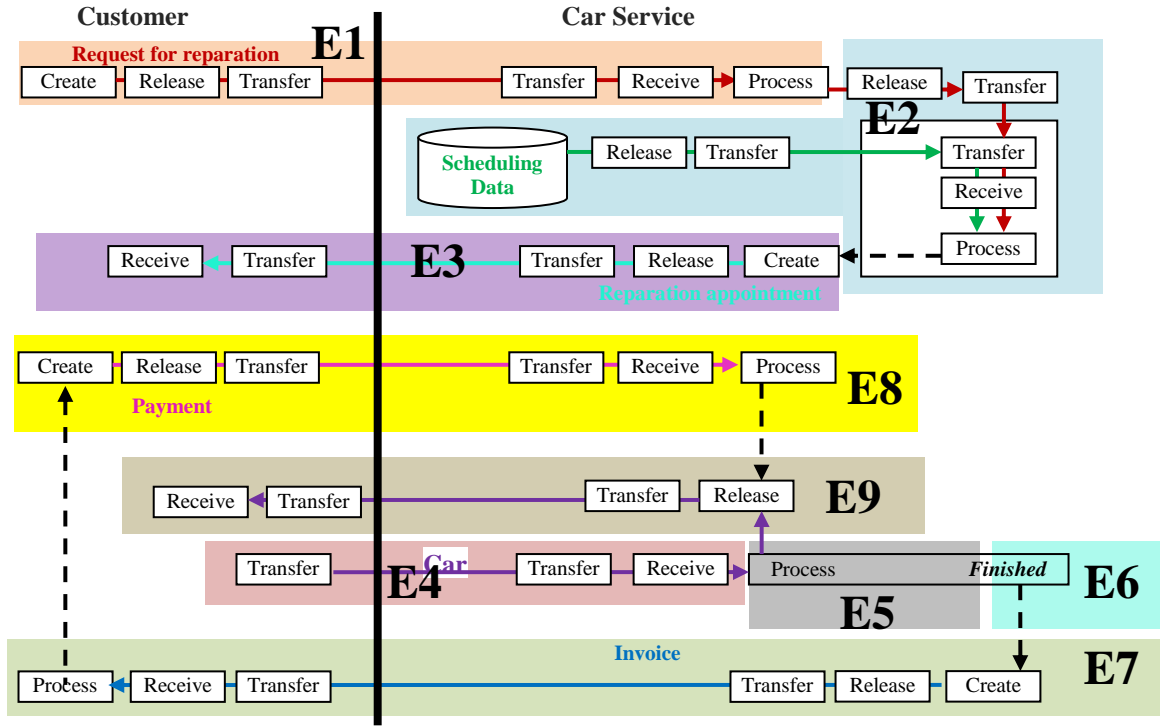

Fig. 19. TM events model

E5: The car is processed.
E6: The car repair is finished.
E7: An invoice is created and sent to the customer.
E8: The customer sends payment to the car service.
E9: The car is released to the customer.
Fig. 20 shows the behavior model.

The given examples in previous sections demonstrate that the TM model with five elementary actions seems to furnish a rich foundation for the notion of action in conceptual system modeling. It also seems that such an approach to express actions can be applied to the study of actions in other fields, such as the semantics for natural languages. The next section provides a sample of this application that will be explored further in the future.

## V. OTHER ACTION-RELATED APPLICATIONS OF THE TM MODEL

Davidson (1967) viewed action sentences in terms of what are called *events* [9]. For example, *Brutus violently stabbed Caesar* is translated (ignoring tense) into a logical expression as follows:

$$(\exists e)(stab(e,Brutus,Caesar) \land violent(e))(\exists e) \ (stab \ (e,Brutus,Caesar) \land violent(e)).$$

This captures the fact that this sentence logically entails that Brutus stabbed Caesar [9]. Fig. 21 shows the TM static representation of *Brutus violently stabbed Caesar*. Fig. 22 shows *Brutus violently stabbed Caesar* in terms of two TM events.

Later, researchers classified verbs into groups: states, activities, accomplishments, achievements, and points. It is to be noted, for example, that verbs of the accomplishments type can be used with the *progressive* tense, whereas verbs of the achievements type cannot.

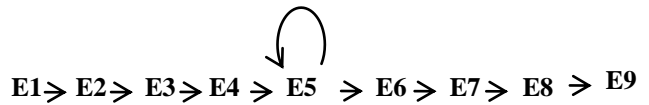

Fig. 20. TM behavior model

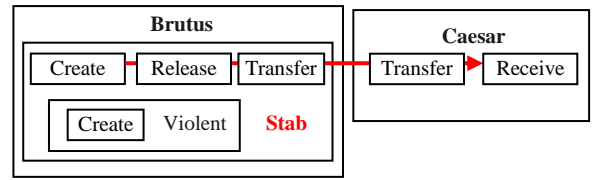

Fig. 21. Static description of *Brutus violently stabbed Caesar*

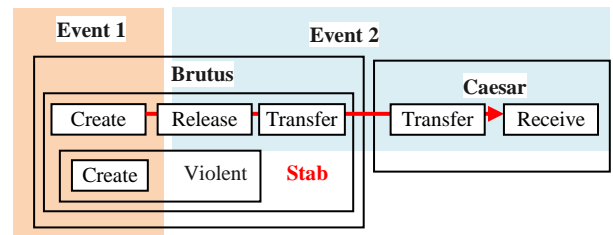

Fig. 22. *Brutus violently stabbed Caesar* as two events

If, for example, I write a letter, the progress is measured in amounts of words. The letter is therefore the incremental theme in *I write a letter* because it defines the *progress* [9]. TM modeling represents this *I write a letter* progression as shown in Fig. 23. The upper left diagram shows the static description, where a letter is created as the result of creating words. The upper right diagram shows the following three events:

E1: A word is created.
E2: A word is processed (i.e., written).
E3: A letter is created.
The lower diagram of Fig. 23 shows that E1 and E2 are repeated until a letter is created.

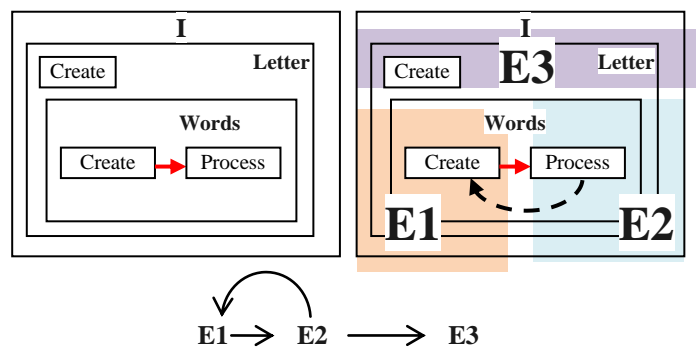

**Fig. 23. Modeling progress in *I write a letter***

## CONCLUSION

In this paper, we focus on the notion of action in the area of conceptual system modeling. In UML and other modeling notation, the meaning of the concept of action is difficult to grasp, and many existing systems involve the notion of action, but in an obscure way. This paper contributes to the establishment of a broader interdisciplinary understanding of the notion of action in conceptual modeling based on a model called a *thinging machine* (TM). The goal of this venture is to improve the process of developing conceptual models by refining basic concepts. The TM approach to the notion of action provides five elementary actions; events are built from these actions when supplemented with time. Accordingly, the conceptual model can start by building a static representation with actions before identifying events and the system's behavior as the chronology of these events. At the end of the paper, we have shown that the same five elementary actions can be used in other fields, such as natural language semantics.

The results in this paper reveal the viability of the TM's five actions in modeling and relate them to other important notions, such as activity, events, and behavior. Further research will involve experimenting with this approach in more examples and using it in various application areas.